\documentclass[pre,amssymb,amsmath,floats,twocolumn,showpacs,superscriptaddress]{revtex4}

\usepackage{epsfig}

\begin{document}

\title{Preferential compactness of networks}

\author{M. J. Alava}
\email{Mikko.Alava@hut.fi}
\affiliation{Helsinki University of Technology, Laboratory of Physics, HUT-02105 Finland}

\author{S. N. Dorogovtsev}
\email{sdo@fyslab.hut.fi, sdorogov@fis.ua.pt}
\affiliation{Helsinki University of Technology, Laboratory of Physics, HUT-02105 Finland}
\affiliation{Departamento de F{\'\i}sica da Universidade de Aveiro, 3810-193 Aveiro, Portugal}
\affiliation{A. F. Ioffe Physico-Technical Institute, 194021
  St. Petersburg, Russia}

\date{}

\begin{abstract}
We introduce evolving networks where new vertices preferentially 
connect to the more central parts of a network. This makes such
 networks compact. Finite networks grown under the preferential compactness mechanism 
have complex architectures, but infinite ones tend towards the opposite,
having rapidly decreasing distributions of connections. 
We present an analytical solution of the problem 
for tree-like networks. 
Our approach links a collective self-optimization mechanism of the emergence of complex network architectures to self-organization mechanisms. 

\end{abstract}


\maketitle

\section{Introduction}\label{s-introduction}


Self-organization versus optimization---these two possible explanations of specific network architectures have been extensively discussed in the last few years. 
The former
mechanism 
is more developed and is a standard explanation for fat-tailed degree distributions in networks in modern physics literature \cite{ba99,baj99,p76}. The optimization concept of the emergence of a complex network structure is more common for engineers and computer scientists \cite{cd99,mg01,vsf02,fkp02}. 
(For more general aspects of self-organization and optimization concepts see Refs. \cite{y25,s55} and Ref.~\cite{mbook83}, respectively.)

Usually these two explanations are seen as mutually contradicting \cite{wgj02,cjw03}. However, the process of optimization of a complex network includes not an external will (a single designer) but numerous agents. Each of these agents (e.g., vertices) solves its ``selfish'', individual optimization problem---optimizes a number of trade-offs---and tries to arrange its connections in the best (for this agent) way. 
Speaking in simple terms, one can even treat optimization of this kind in terms of self-organization and vice versa. 

To be more concrete, we consider a more narrow class of models. 
The simplest one, a model of a growing tree, where each new vertex becomes attached to one of existing vertices selected in an optimal way, was proposed by Fabrikant, Koutsoupias, and Papadimitriou, Ref.~\cite{fkp02} (for the detailed study of the model, see Ref.~\cite{bbbcr03}, see also Ref.~\cite{bbcsk04}). In the FKP model vertices have random geographical coordinates in a restricted area. A new vertex is attached to one of vertices chosen to minimize a linear combination of (i) the resulting shortest-network-path distance between the new vertex {\em and the root} 
of the tree 
 and (ii) the Euclidean length of the new connection.  

The FKP model exploits the  
competition of two objectives: a desire to connect to the center of the network, which produces compact networks, and a desire to have a physically short, cheap connection. 
The geographical coordinates of vertices are taken to be random, as well as, in fact, the
Euclidean distances between vertices. So, one may roughly say that in this scheme, 
the combination of a length to the center of a network and a random number is optimized. Consequently, the competition is actually between compactness and randomness. 

In this paper we study the interplay between these two tendencies to compactness and to randomness as the mechanism of generation of complex networks. 
We formulate the problem in terms of preferential attachment and so link the optimization models to self-organization ones. 

We introduce the following model of a growing network (tree): 

\begin{itemize}

\item[(1)] 
At each time step, a vertex is added to the network. 

\item[(2)] 
This vertex is attached to a vertex chosen with probability proportional to a function $r(\ell)$ of its distance $\ell$ from the root.  

\end{itemize}
The preference function $r(\ell)$ determines the structure of the network. As is natural, if $r(\ell)=\text{const}$, we arrive at the random attachment growth and at an exponential degree distribution. We show that if $r(\ell)$ is rapidly decreasing, the resulting degree distribution of the finite network is of a complex form with several distinct parts. On the other hand, we find that in the infinite network limit, the degree distribution is a rapidly decreasing function. 
Note the difference of our approach from that of Ref.~\cite{yjb01} where preferential attachment of new vertices was determined by geographical closeness (Euclidean distance), see also Refs. \cite{a02,rcbh02,xs02}. 

\begin{figure}
\epsfxsize=74mm
\epsffile{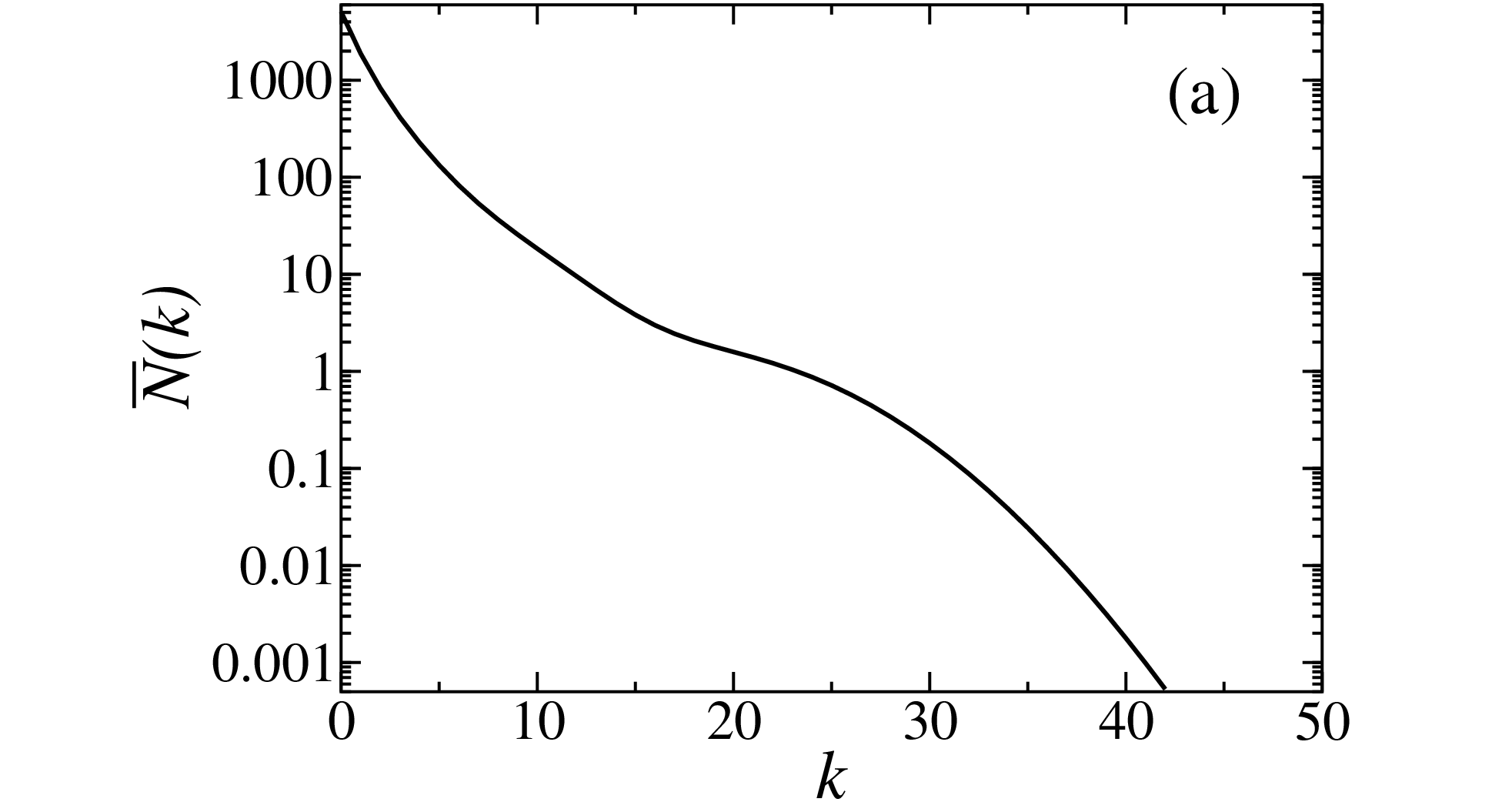}
\epsfxsize=60mm
\epsffile{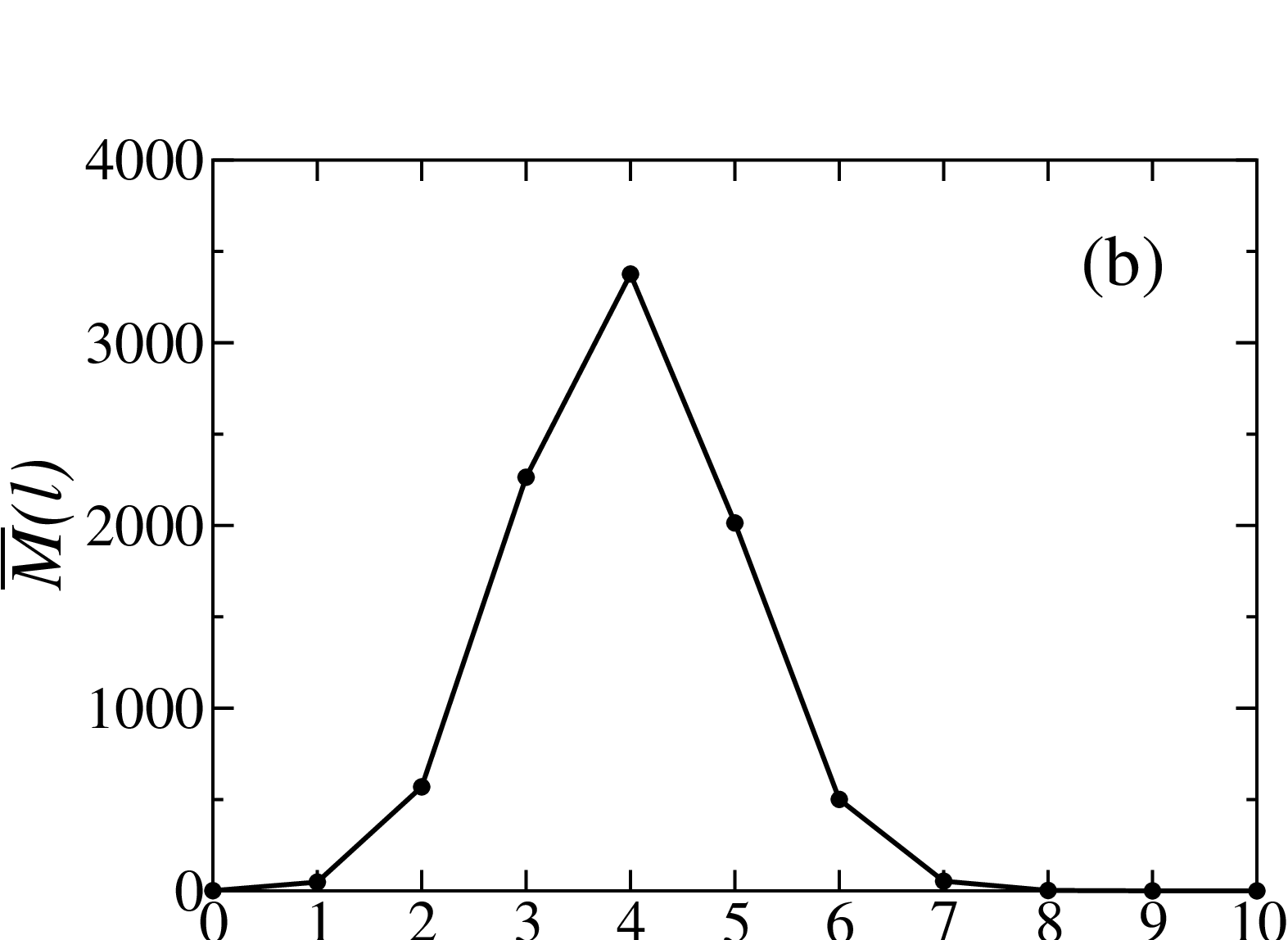}
\caption{
(a) The degree distribution of the network growing under the mechanism of preferential compactness with the preference function $r(\ell)=x^\ell$. 
The value of the parameter $x$ is 0.5, the size of the network is $t=8825$. 
Equation (\protect\ref{e10}) gives $\tau=47.72$ for these values. 
The width of the plateau is $x\tau=24$, the height of the plateau 
$1/x=2$. 
(b) The distribution $\overline{M}(\ell)$ in this network 
(the mean number of vertices at a distance $\ell$ from the root). Notice that the number of the nearest neighbors of the root, $\overline{M}(\ell=1)=\tau$, here is a vanishingly small fraction of the total number of vertices. 
}
\label{f1}
\end{figure}

\begin{figure}
\epsfxsize=61mm
\epsffile{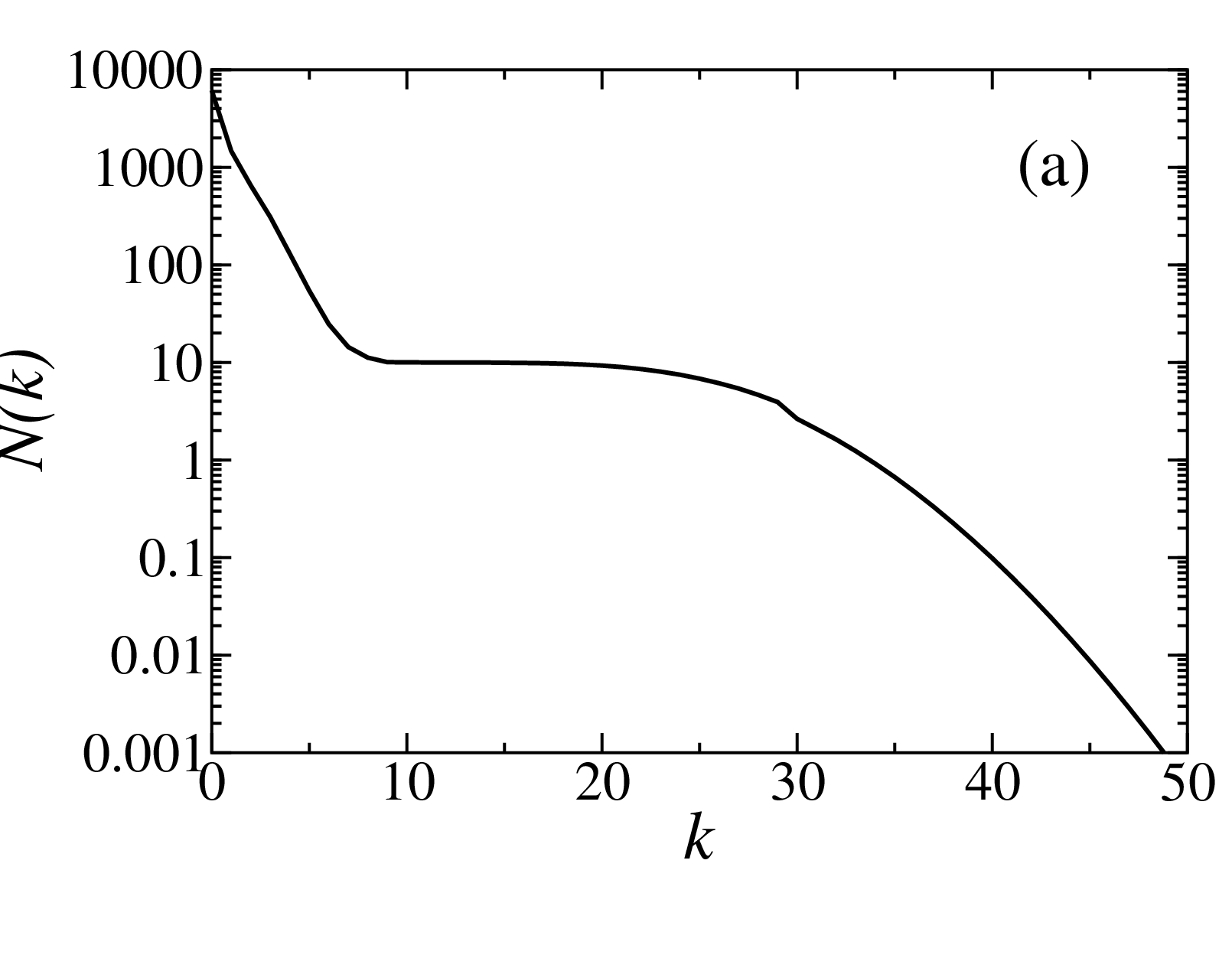}
\epsfxsize=60mm
\epsffile{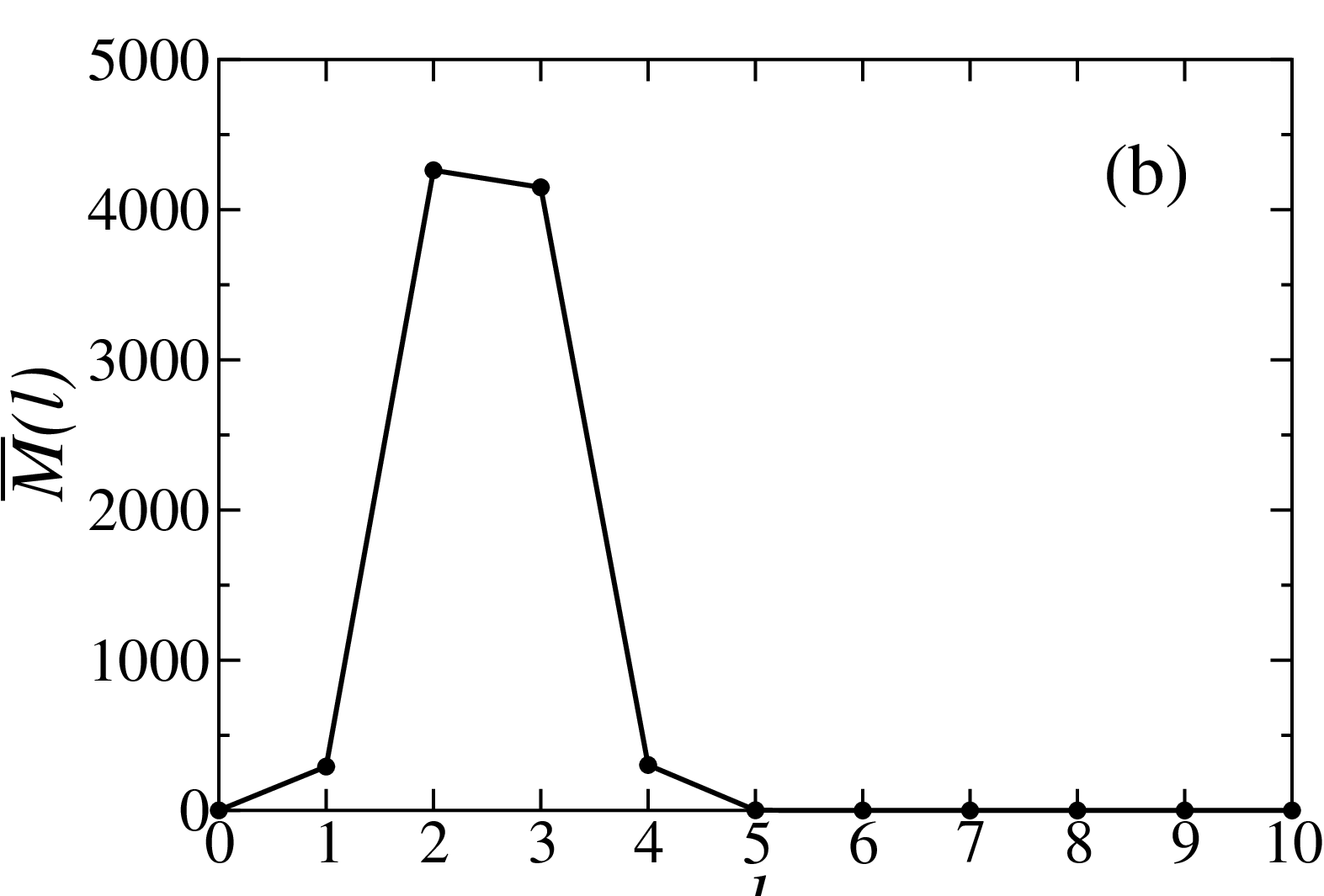}
\caption{
(a) The same as in Fig.~\protect\ref{f1} but $x=0.1$, $t=9005$.  
Equation (\protect\ref{e10}) gives for these values $\tau=292$. 
The width of the plateau is $x\tau=29$, the height of the plateau 
is $1/x=10$.
(b) The distribution $\overline{M}(\ell)$ in this network. 
Note that $\overline{M}(\ell=1)$ is noticeable with these values 
of $x$ and $t$. 
}
\label{f2}
\end{figure}

\begin{figure}
\epsfxsize=61mm
\epsffile{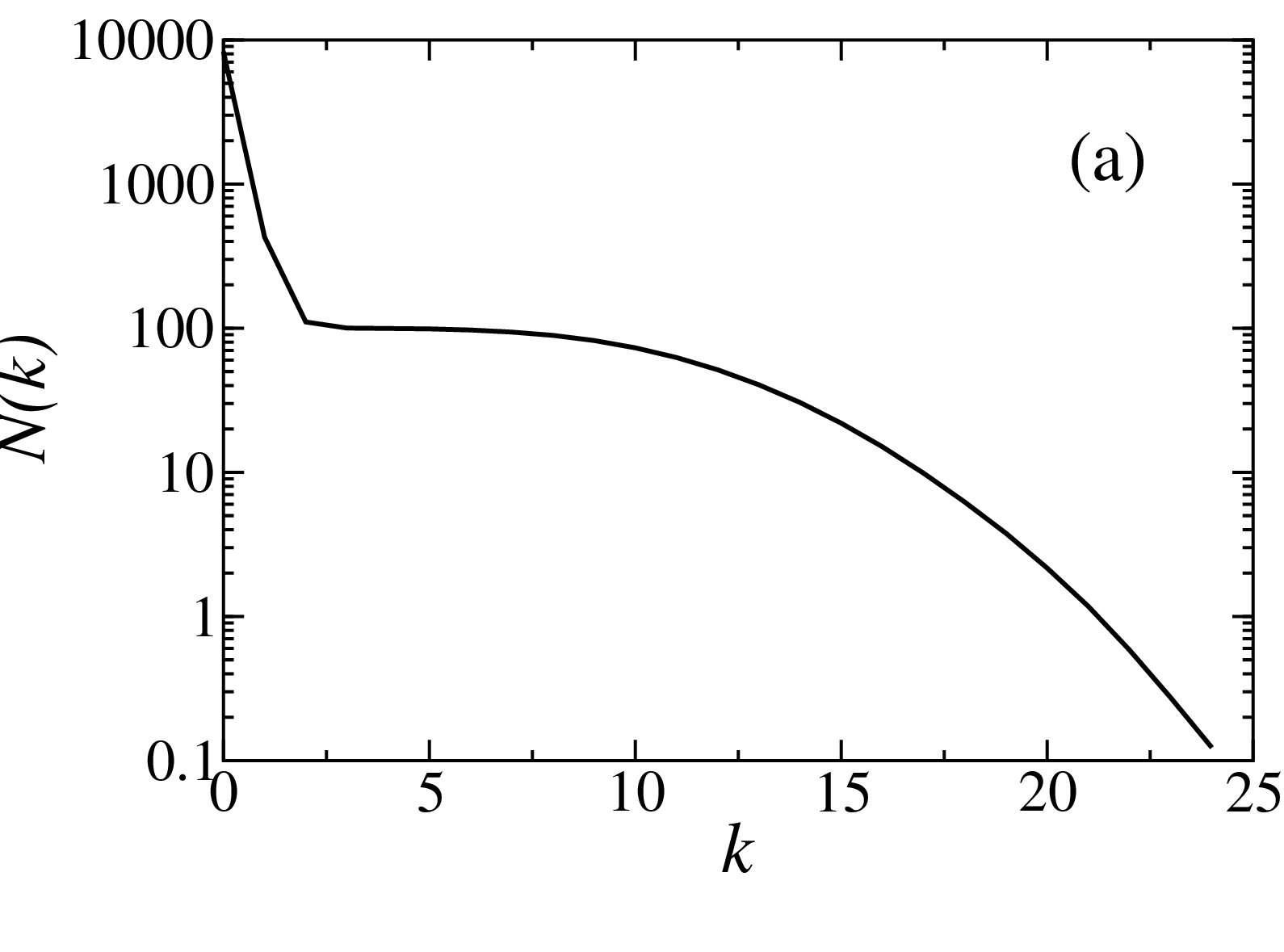}
\epsfxsize=60mm
\epsffile{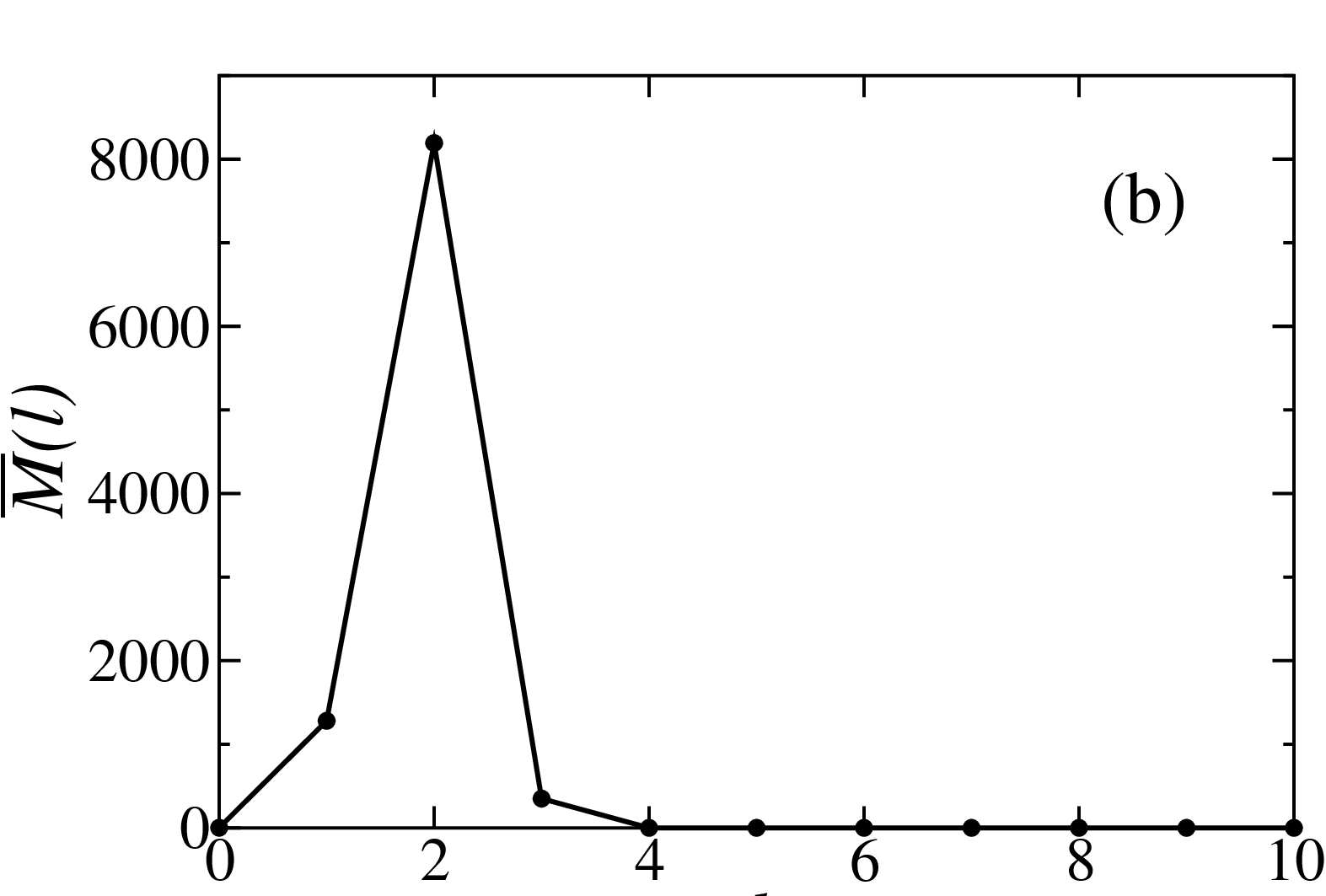}
\caption{
(a) The same as in Figs~\protect\ref{f1} and \protect\ref{f2} but $x=0.01$, $t=9821$.  
Equation (\protect\ref{e10}) gives for these values $\tau=1280$. 
The width of the plateau is $x\tau=13$, the height of the plateau 
is $1/x = 100$.
(b) The distribution $\overline{M}(\ell)$ in this network. 
$\overline{M}(\ell=1)$ is noticeable with these values of $x$ and $t$, which results in the pronounced plateau in the degree distribution.  
}
\label{f3}
\end{figure}

Our results, typical resulting degree distributions in finite networks are shown in Figs.~\ref{f1}(a), \ref{f2}(a), and \ref{f3}(a). Note that the numbers of vertices in these examples far exceed typical sizes of networks in simulations of network optimization (see, e.g., Refs.~\cite{vsf02,cbmr04}) and in most of the studied real networks. 
In this sense, the model provides complex degree distributions even in ``large'' networks. 

In the present paper we explain the nature of this behavior of the degree distribution in networks of this kind. We mean the combination of a complex form in the finite networks, and a ``trivial'' one in the thermodynamic limit.  
Furthermore, we show that one can essentially extend the range of the observation of the complex degree distributions by passing to more realistic models. 

Note that a tree structure was assumed only to simplify analytical calculations (it is easy to find intervertex distances in trees). The basic idea of the model, that is, the specific preferential linking that makes a network more compact, may be also applied to networks with loops. 


\section{Evolution equations}\label{s-equation}

The tree structure of the network 
simplifies the problem. 
It is convenient to treat edges as directed, outgoing from newer vertices. So, we consider the statistics of in-degree $k$ of vertices, and the quantity of interest is the average number $\overline{N}(k,\ell,t)$ of vertices of in-degree $k$ at a distance $\ell$ from the root at time $t$. (``Time'' is the current number vertices in the network.) Here the average means averaging over the entire ensemble of networks at time $t$. This is a set of networks which can emerge at time $t$ as a result of the evolution of the model with statistical weights determined by the rules of this evolution. 

We use an approach, standard in networks with preferential 
linking \cite{krl00,dms00}. 
Let $\overline{M}(\ell,t)$ be the average number of vertices at a distance $\ell$ from the root at time $t$. The mean in-degree of the root vertex is equal to the mean number of the nearest neighbors of the root, $\overline{k}(\ell=0,t) = \overline{M}(\ell=1,t)$. $\overline{M}(\ell,t) = \sum_{k=0}\overline{N}(k,\ell,t)$. Note that in this tree, the out-degree of any non-root vertex is $1$, so the total degree of a vertex is $k+1$. 

One can see that the evolution of $\overline{M}(\ell,t)$ is described by the equation: 

\begin{equation}
\overline{M}(\ell,t+1) = 
\overline{M}(\ell,t) + \frac{r(\ell-1)}{\Sigma(t)}\overline{M}(\ell-1,t)
\, ,
\label{e1}
\end{equation} 
where 

\begin{equation}
\Sigma(t) = \sum_{\ell=0}r(\ell)\overline{M}(\ell,t)
\, .
\label{e2}
\end{equation} 
An initial condition is $\overline{M}(\ell,t=1) = \delta_{\ell,0}$, where $\delta_{\ell,0}$ is the Kronecker symbol. 
That is, the growth starts from a single vertex---the root, and $t$ is exactly the number of vertices in the net.  
As is natural, Eqs.~(\ref{e1}) and (\ref{e2}) guarantee that 
$\overline{M}(\ell=0,t\geq1)=1$. The solution $\overline{M}(\ell,t)$ must satisfy the condition 

\begin{equation}
\sum_{\ell=0}\overline{M}(\ell,t) = t 
\, .
\label{e3}
\end{equation} 
It is important that Eq.~(\ref{e1}) does not contain  
$\overline{N}(k,\ell,t)$. The evolution of $\overline{N}(k,\ell,t)$ is governed by the equation: 

\begin{eqnarray}
& & \overline{N}(k,\ell,t+1) = 
\overline{N}(k,\ell,t) 
\nonumber
\\[5pt]
& &
+ \frac{r(\ell)}{\Sigma(t)}
[\overline{N}(k-1,\ell,t) - \overline{N}(k,\ell,t)] 
\nonumber
\\[5pt]
& &
+ 
\frac{r(\ell-1)}{\Sigma(t)}\overline{M}(\ell-1,t)\delta_{k,0}
\, \vspace{10pt}. 
\label{e4}
\end{eqnarray} 
The second term on the right-hand side of this equation accounts for the attachment of new vertices to vertices of an $\ell$th shell. 
The third term is due to emergence of extra leaves in the $\ell$th shell when new vertices become attached to vertices at distance $\ell-1$ from the root.

\section{Calculations}\label{s-calculations}

We must first find a solution $\overline{M}(\ell,t)$ of Eq.~(\ref{e1}), substitute this solution to Eq.~(\ref{e4}), and then obtain the result $\overline{N}(k,\ell,t)$. Note that the solutions are non-stationary.  

We consider a large $t$ asymptotic behavior, so that the differences 
$\overline{M}(\ell,t+1)-\overline{M}(\ell,t)$ in Eq.~(\ref{e1}) and $\overline{N}(k,\ell,t+1)-\overline{N}(k,\ell,t)$ in Eq.~(\ref{e4}) can be substituted by the corresponding time derivatives. 

The direct way to solve the problem is as follows: 
(i) find the solution of Eq.~(\ref{e1}) in terms of yet unknown function $\Sigma(t)$, i.e., $\overline{M}(\ell,t, \Sigma(t))$; 
(ii) substitute this solution into Eq.~(\ref{e3}) [or Eq.~(\ref{e2})] and obtain $\Sigma(t)$;  
(iii) with this $\Sigma(t)$, obtain the solution $\overline{M}(\ell,t)$, 
(iv) solve Eq.~(\ref{e4}) with the known $\Sigma(t)$ and $\overline{M}(\ell,t)$. 

Technically, it is convenient to use the mean number of the nearest neighbors of the root, $M(\ell=1,t) \equiv \tau(t)$, instead of $\Sigma(t)$. 
One can see that $\tau(t)$ and $\Sigma(t)$ are related to each other in a simple way. 
Without restriction of generality, we set $r(0)=1$, so Eq.~(\ref{e1}) readily gives the relation:   

\begin{equation}
\frac{\partial \overline{M}(\ell=1,t)}{\partial t} = \frac{d \tau(t)}{dt} = \frac{1}{\Sigma(t)} 
\, .
\label{e5}
\end{equation} 
Then, for $\overline{M}_\ell(\tau(t)) \equiv \overline{M}(\ell,t)$ and 
$\overline{N}_{k,\ell}(\tau(t)) \equiv \overline{N}(k,\ell,t)$, we have 
the equations: 

\begin{equation}
\frac{\partial \overline{M}_\ell(\tau)}{\partial \tau} = r(\ell-1)\overline{M}_{\ell-1}(\tau) 
\, 
\label{e6}
\end{equation} 
and 

\begin{eqnarray}
\frac{\partial \overline{N}_{k,\ell}(\tau)}{\partial \tau} 
& = &
r(\ell)[\overline{N}_{k-1,\ell}(\tau) - \overline{N}_{k,\ell}(\tau)] 
\nonumber
\\[5pt]
& &
+
r(\ell-1)\overline{M}_{\ell-1}(\tau)\delta_{k,0}
\, . 
\label{e7}
\end{eqnarray}  
The solution of Eq.~(\ref{e6}) is 

\begin{equation}
\overline{M}_\ell(\tau) = \frac{\tau^\ell}{\ell!}\, \prod_{i=0}^{\ell-1}r(i) 
\, . 
\label{e8}
\end{equation} 
Let us now, for simplicity, use a concrete form of the preference function, $r(\ell) = x^\ell$, where $0<x<1$. 
[$x=1$ corresponds to random (indifferent) attachment; $x>1$ with large enough $x$ results in chain-like structures.] In this case, the solution (\ref{e8})
is ``Poisson''-like: 

\begin{equation}
\overline{M}_\ell(\tau) = \frac{1}{\ell!}\,x^{\ell(\ell-1)/2}\, \tau^\ell 
\, , 
\label{e10}
\end{equation} 
with $\tau$ satisfying the equation: 

\begin{equation}
1+ \sum_{\ell=1}\frac{1}{\ell!}\,x^{\ell(\ell-1)/2}\, \tau^\ell = t 
\, . 
\label{e11}
\end{equation} 
The final formulas will express the dynamics of the model in terms of $\tau$, and the time dependence $\tau(t)$ is the result of the solution of the transcendental equation (\ref{e11}).

The asymptotics of the resulting dependence of the number of the nearest neighbors of the root, 

\begin{equation}
\tau(t) \cong e^{-1} 
\sqrt{\frac{2x}{\ln (1/x)}}\, \sqrt{\ln t}\, \exp\left[\sqrt{2\ln (1/x) \ln t}\right] 
\, ,  
\label{e12}
\end{equation} 
can be obtained in situations where the peak of $\overline{M}_\ell$ is well separated 
from $\ell=0$,  
so the saddle point method is applicable. Actually, for the validity of the asymptotic form (\ref{e12}), the point of the maximum of $\overline{M}_\ell$, $\ell_{\text{max}} \sim \overline{\ell}$  must essentially exceed the width $\delta \ell$ of this distribution. ($\overline{\ell}$ is the average separation of a vertex from the root.) 
In its turn, the dispersion $\delta \ell$ must be essentially larger than $1$, which is not the case for physically reasonable values of parameters. 

Expression (\ref{e12}) can be rewritten in the following form

\begin{equation}
\tau(t) \sim \sqrt{\ln t}\, t^{\sqrt{2\ln(1/x)}/\sqrt{\ln t}}
\, .   
\label{e13}
\end{equation} 
That is, $\tau$ grows slower than any power of $t$ but at finite $t$, looks rather as a ``power law'' with a slowly varying exponent 
(a ``multifractal'' appearance). 

The characteristics $\ell_{\text{max}}$ (a typical distance from the root) and $\delta\ell$ of the distribution $\overline{M}_\ell$ have asymptotics: 

\begin{equation}
\ell_{\text{max}} 
\sim \frac{\ln\tau}{\ln(1/x)} 
\,   
\label{e14}
\end{equation} 
and 
\begin{equation}
\delta\ell \sim 2\left( \ln(1/x) + \frac{1}{\ell_{\text{max}}} - \frac{1}{2\ell_{\text{max}}^2} \right)^{-1/2}
\, .  
\label{e15}
\end{equation} 
So that, very roughly, $\ell_{\text{max}} \sim \sqrt{2\ln t/\ln(1/x)}$. 

Unfortunately, these compact asymptotic formulas become precise only in astronomically large networks. For ``physically reasonable'' values of parameters (see Figs \ref{f1}, \ref{f2}, and \ref{f3}), it is simpler to numerically obtain $\tau(t)$ from Eq.~(\ref{e11}) than to use cumbersome next order asymptotic formulas. 

The solution of Eq.~(\ref{e7}) with the substituted $\overline{M}_{\ell}(\tau)$, formula (\ref{e10}), is 

\begin{eqnarray} 
\overline{N}_{k,\ell\geq1}(\tau) & = &
\frac{(-1)^{\ell+1}}{x^{\ell(\ell+1)/2}} 
\left[  
\sum_{i=0}^{\ell-1} (-1)^i {k+\ell-1-i \choose \ell-1-i}\frac{(x^\ell\tau)^i}{i!}\right. 
\nonumber
\\[5pt]
& &
\!\!\!\!\!\!\!\!\!
\left.- e^{-x^\ell\tau} \sum_{j=0}^{k} {k+\ell-1-j \choose \ell-1}\frac{(x^\ell\tau)^j}{j!}
\right]
\, .  
\label{e16}
\end{eqnarray} 

Expressions (\ref{e10}) and (\ref{e16}) with $\tau(t)$ obtained from Eq.~(\ref{e11}) give the complete solution of the problem. 
It is useful to write out  particular cases of the cumbersome expression (\ref{e16}): for the nearest neighbors of the root, we have 

\begin{eqnarray} 
\overline{N}_{k,1}(\tau) 
& = & 
\frac{1}{x} 
\left[  
1 - e^{-x\tau}\sum_{i=0}^{k}\frac{(x\tau)^i}{i!}
\right]
\nonumber
\\[5pt]
& = &
\frac{1}{x}\,e^{-x\tau}\sum_{i=k+1}^{\infty}\frac{(x\tau)^i}{i!}
\, ,  
\label{e17}
\end{eqnarray} 
for the second-nearest neighbors, we have 

\begin{eqnarray} 
\overline{N}_{k,2}(\tau) 
& = & 
\frac{(-1)}{x^3} 
\left[  
k+1 - x^2\tau 
\right. 
\nonumber
\\[5pt]
& &
\left.
- e^{-x^2\tau} \sum_{i=0}^{k}(k+1-i)\frac{(x^2\tau)^i}{i!}
\right]
\, ,   
\label{e18}
\end{eqnarray} 
and so on.

\section{Results and discussion}\label{s-results}

Typical $\overline{N}(k,t) = \sum_{\ell}\overline{N}(k,\ell,t)$, obtained by the direct summation of expression (\ref{e16}), are shown in Figs~\ref{f1}(a), \ref{f2}(a), and \ref{f3}(a). [A degree distribution is $P(k,t) = \overline{N}(k,t)/t$.] 
The corresponding mean numbers of vertices at a distance $\ell$ from the root are shown in Figs~\ref{f1}(b), \ref{f2}(b), and \ref{f3}(b). 
  
The degree distributions consist of three parts: there is a plateau between two regions of rapid decrease at small and large degrees. Note that the main fraction of vertices are leaves (vertices of degree 1). 

What is the nature of this complex form of the degree distribution, i.e., the nature of the plateau? The inspection of expression (\ref{e16}) shows that $\overline{N}(k,\ell,t)$ rapidly decreases with $k$ at each $\ell>1$. [Check, e.g., in the particular case $\ell=2$, relation (\ref{e18}).] On the other hand, expression (\ref{e17}) ($\overline{N}_{k,\ell=1}$) shows that the plateau is determined by the contribution of the nearest neighbors of the root. Recall that the total number of these vertices is $\tau(t)$. One can see from expression (\ref{e17}) that $\overline{N}_{k,\ell=1}$ is close to $1/x$ as $k \lesssim x\tau$. Above this value of in-degree, $\overline{N}_{k,\ell=1}$ is a rapidly decreasing 
dependence. 
Note that $\sum_k \overline{N}_{k,\ell=1}(t) = \overline{M}_{\ell=1}(t) = \tau(t)$. 
Figs.~\ref{f1}(a), \ref{f2}(a), and \ref{f3}(a) demonstrate that indeed the width of the plateau is $x\tau$, and its height is $1/x$.  

So, for the presence of the plateau in the degree distribution two conditions must be simultaneously fulfilled: (i) $x\tau$ must be much greater than the mean in-degree of the network, i.e., much greater than 1; (ii) $1/x$ must be large enough. 
Figs.~\ref{f2}(a) and \ref{f3}(a) show situations in finite networks where these conditions are satisfied. One can check that if the size of the network tends to infinity, the relative number of the nearest neighbors of the root approaches zero, $\tau/t\to 0$, and the plateau disappears. 

Thus, it is the nearest neighbors of the root that make the degree distribution of the network complex. The decrease of the fraction of these vertices in large networks makes the plateau unobservable and so trivializes the form of the degree distribution. We believe that this is a rather generic feature of networks of this kind. We mean networks, generated by optimization related algorithms, where a function of an intervertex distance is used in optimization (or for selection a vertex for attachment). 
Quite similarly to what we observed, in the FKP model, the ``complex'', power-law part of the degree distribution disappears in the thermodynamic limit and vertices are almost surely leaves \cite{bbcsk04}. 

One might consider this trivialization in the thermodynamic limit as a defect of these constructions. However, most of real networks are not so large. Empirical studies and simulations are mostly made for networks smaller than the networks in Figs~\ref{f1}--\ref{f3}. Moreover, the typical sizes of networks in simulations of optimization mechanisms are 200, 100, and 50 vertices (see, e.g., Refs.~\cite{vsf02,cbmr04}). In this sense, we observe complex architectures in quite large networks. 


\begin{figure}
\epsfxsize=61mm
\epsffile{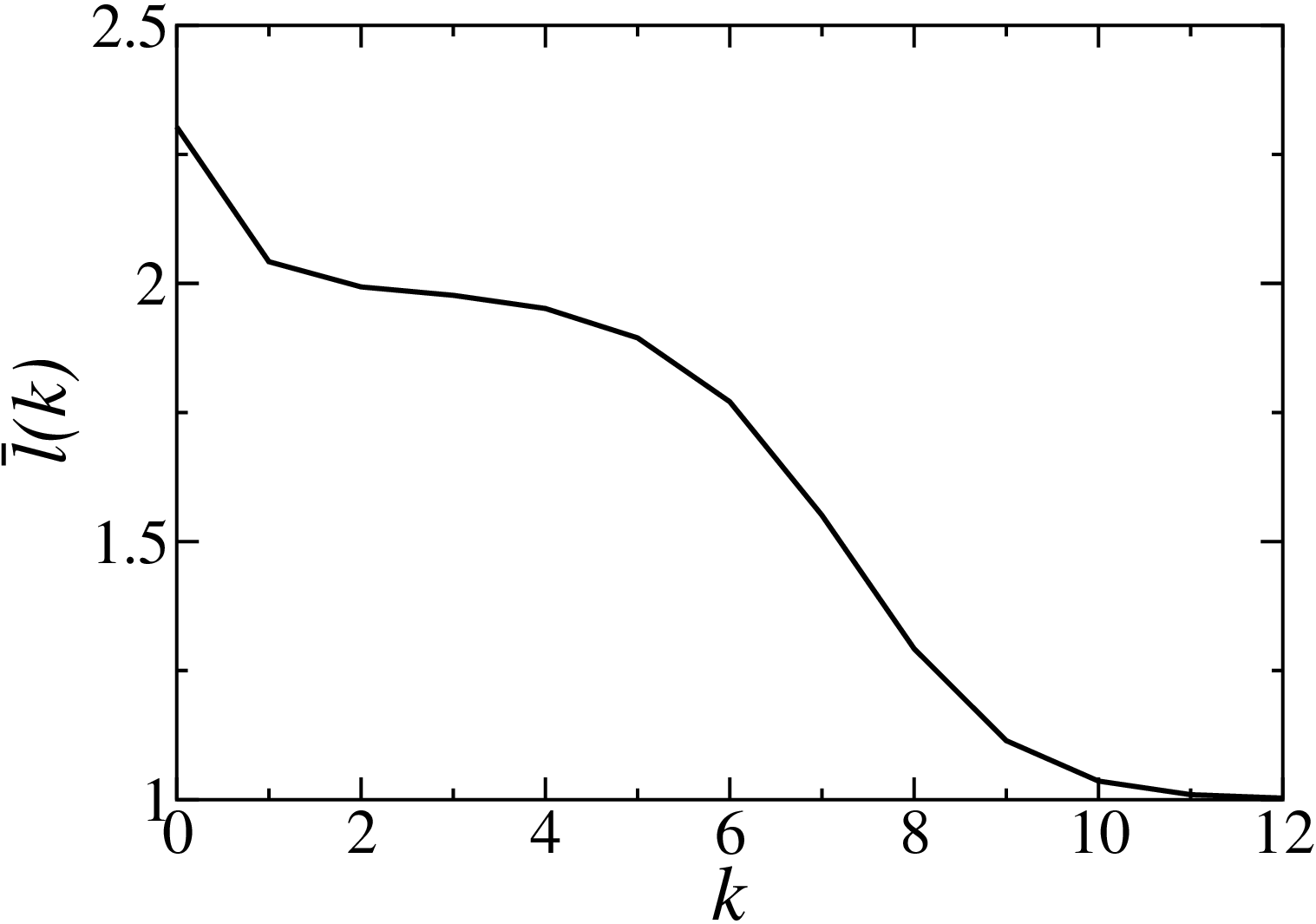}
\caption{
The mean separation of a vertex from the root as a function of the degree of this vertex. $x=0.1$, $t=20142$, and so $\tau=400$. The dependence exponentially approaches 1 as $k$ increases. 
}
\label{f4}
\end{figure}


The diminishing of the fraction of the nearest neighbors of the root, $\tau(t)/t$, in the network, which we consider in this paper, is a direct consequence of the narrow distribution of vertices over the distance $\ell$ from the root [see Figs~\ref{f1}(b), \ref{f2}(b), and \ref{f3}(b)]. As the network grows, the mean distance of a vertex from the root, $\overline{\ell}$ increases as $\sqrt{\ln t}$ (which is meanwhile slower than a standard logarithmic dependence  typical for networks with the small-world effect). Simultaneously, the relative width of the distribution decreases [see relation (\ref{e14})], and approaches zero in the large network limit. it is important that this behavior of the relative width is practically generic for networks with the small-world effect. In other words, in infinitely large networks with the small-world effect, all vertices are almost surely mutually equidistant.     
So, in the large networks, even rapidly, exponentially decreasing preference function $r(\ell) \propto x^\ell$ leads to attachments mostly to vertices at the distance $\overline{\ell}$ from the root. Consequently, in the large networks, the nearest neighbors of the root have few chances to attract new connections

In another form, the result of this evolution is shown in Fig.~\ref{f4}: the average separation $\overline{\ell}(k)$ of a vertex from the root as a function of the degree of this vertex in a typical situation. 
This figure demonstrates that most vertices have the same separation from the root. 
Note that the complementary characteristic, the mean in-degree of a vertex at distance $\ell$ from the root shows a quite different behavior: 

\begin{equation}
\overline{k}(\ell,t) = \frac{M(\ell+1,t)}{M(\ell,t)} = \frac{x^\ell \tau(t)}{\ell+1}
\, 
\label{e19}
\end{equation}             
[we have used relation (\ref{e10})]. 

A number of modifications of the model are possible:  

(1) An analytical solution has been obtained for the tree network, but in principle, one may also allow connections between already existing vertices using the same preference function.

(2) One may introduce the variation of the preference function with the network growth, e.g., use $r(\ell,t) = x^\ell(t)$. This allows us to increase the range of network sizes, where the degree distribution has a complex form, but the model becomes ugly. 

(3) One may introduce an additional finite probability that a new vertex is attached to the root of the network. This leads to a finite fraction of the nearest neighbors of the root [$\tau(t) \propto t$, the condensation of edges on the root], and so to the presence of the plateau in the degree distribution even in the infinite network. 

(4) If, nevertheless, we want to avoid the condensation of edges, we may, for example, modify the rules of the model in the following way. We introduce equiprobable attachment of a new vertex to the root or to any of its nearest 
neighbors. 
In other words, the ``center'' of the network consists now of the root and its nearest neighbors.  

Let us discuss the last possibility in more detail. The evolution of the average degree of the root, $\tau(t)$, is described by the equation: 
\begin{equation} 
\frac{d\tau}{dt} = p\frac{1}{1+\tau} 
\, ,
\label{e20}
\end{equation}  
where $p$ is the probability that a new vertex becomes attached to the ``center'' of the network. The initial condition is $\tau(t=0)=0$, so the solution of the equation is $\tau \cong \sqrt{2pt}$. 

It is convenient to use $\tau$ as a new ``time'' variable, instead of $t$. In this case, the number of vertices in the ``center'' of the network per ``time'' step increases by 1. On the other hand, the number of vertices, which become attached to the ``center'' vertices per time step, increases with $\tau$ as $dt/d\tau \cong \tau/p$. So, the equation for the average number of the vertices of in-degree $k$ in the ``center'' of the network is of the form: 
\begin{eqnarray} 
&&
\overline{N}(k,\tau+1) = \overline{N}(k,\tau) + \delta_{k,0} 
\nonumber
\\[5pt]
&&
+ 
\frac{\tau}{p}\,
\left[ 
\frac{\overline{N}(k-1,\tau)}{\tau} - \frac{\overline{N}(k,\tau)}{\tau}
\right]
\, .
\label{e21}
\end{eqnarray}  
The solution of this equation, $\overline{N}(k,\tau)$, has a kink-like form: 
$\overline{N}(k<\tau/p,\tau) \cong p$, $\overline{N}(k>\tau/p,\tau) \cong 0$. 
So, the width of the plateau is $p\tau$. Recall that in the original model, we had the plateau width $x\tau$ with an extremely slowly growing $\tau(t)$.  
In contrast, in the present situation, the number of vertices in the ``center'' of the network, $\tau(t)$, and the plateau width rapidly grow as $\sqrt{t}$. 
This allows one to observe highly connected vertices in larger networks than in the original net.  
Note, however, that in the present case, in average, there is no more than one vertex of each degree in the plateau part of the distribution.

\section{Summary}\label{s-summary}

We have proposed a class of networks exhibiting the mechanism of preferential compactness. 
New vertices prefer to be attached to the central part of a net, which favours compact structures. 

Our approach links optimization concepts of complex networks to the self-organization mechanism.  
These networks, while growing, self-organize into compact structures, whose linear sizes are determined by the assumed preference functions. These thus
fix a balance between tendencies to compactness and to randomness in the network growth. 

As an example, we have studied a tree-like growing compact network (the mean intervertex distance grows slower than the logarithm of the number of vertices). 
We have found that the degree distribution of the network has a complex 
(non-power-law, in our particular case) form only in the finite networks. The degree distribution trivializes in the large network limit, where this distribution consists only of a rapidly decreasing part. We have shown that this ``trivialization'' of the network structure is practically unavoidable if an intervertex distance is used in the optimization (or preferential linking) process. However, in our networks, the infinite network limit is approached slowly. 
So, the structures of even reasonably large (but 
still 
finite) networks, generated in this way, are complex. 
The relevant network sizes, where a complex distribution of connections is observable, turn out to be typical for many real and simulated networks. 
Comparison with the FKP model shows that the observed behavior should be common for networks whose evolution is directly determined by separation between their vertices. 
 


In summary, we presented a new effective approach to what one can call the ``collective self-optimization'' of \vspace{-13pt}networks.

\begin{acknowledgments}

MJA is grateful to the Center of Excellence program
of the Academy of Finland for support. SD would like
to acknowledge the hospitality of the Helsinki University
of Technology. A part of the work was made when one of the authors (SD) attended the Exystence Thematic Institute on Networks and Risks (Collegium Budapest), and SD thanks A.~Vespignani for a useful discussion in Budapest.  
SD was partially supported by projects POCTI/FAT/46241/2002 and POCTI/MAT/46176/2002.

\end{acknowledgments}


\begin{thebibliography}{10} 

\bibitem{ba99}
A.-L.~Barab{\'a}si and R.~Albert, 
Emergence of scaling in complex networks, 
Science {\bf 286}, 509 (1999). 

\bibitem{baj99} 
A.-L.~Barab{\'a}si, R. Albert, and H. Jeong, 
Mean-field theory for scale-free random networks, 
Physica A {\bf 272}, 173 (1999). 

\bibitem{p76}
D.J.~de~S.~Price, 
A general theory of bibliometric and other cumulative
advantage processes, 
J.~Amer. Soc. Inform Sci. {\bf 27}, 292 (1976).

\bibitem{cd99}
J.M.~Carlson and J.C.~Doyle, 
Highly Optimized Tolerance: A mechanism for power laws in designed systems, 
Phys. Rev. E {\bf 60}, 1412 (1999); 
Highly Optimized Tolerance: Robustness and design in complex systems,  
Phys. Rev. Lett. {\bf 84}, 2529 (2000); 
M.~Newman, The power of design,  
Nature {\bf 405}, 412 (2000).  

\bibitem{mg01}
N.~Mathias and V.~Gopal, 
Small-worlds: How and why, 
Phys. Rev. E {\bf 63} 021117 (2001). 

\bibitem{vsf02}
S.~Valverde, R.V.~Sol\'e, and R.~Ferrer i Cancho, 
Scale-free networks from optimal design, 
Europhys. Lett. {\bf 60}, 512 (2002). 

\bibitem{fkp02}
A.~Fabrikant, E.~Koutsoupias, and C.H.~Papadimitriou, 
Heuristically optimized trade-offs: A new paradigm for power laws in the Internet, 
LNCS {\bf 2380}, 110 (2003). 

\bibitem{y25}
G.U.~Yule, 
A mathematical theory of evolution, based on the conclusions of 
Dr.~J.C.~Willis, 
Phil. Trans. Royal Soc. London B {\bf 213}, 21 (1925); 
J.C.~Willis, 
{\em Age and Area} (Cambridge University Press, Cambridge, 1922). 

\bibitem{s55}
H.A.~Simon, 
On a class of skew distribution functions, 
Biometrika {\bf 42}, 425 (1955).

\bibitem{mbook83} 
B.B.~Mandelbrot, 
{\em The Fractal Geometry of Nature} (Freeman, New York, 1983). 

\bibitem{wgj02}
W.~Willinger, R.~Govindan, S.~Jamin, V.~Paxson, and S.~Shenker, 
Scaling phenomena in the Internet: Critically examining criticality, 
PNAS {\bf 99}, 2573 (2002). 

\bibitem{cjw03}
H.~Chang, S.~Jamin, and W.~Willinger, 
Internet connectivity at the AS-level: 
An optimization-driven modeling approach,  
Extended Technical Report CSE-TR-475-03, EECS Department, University of Michigan, 2003, http://www.eecs.umich.edu/techreports/cse/2003/CSE-TR-475-03.pdf. 

\bibitem{bbbcr03} 
N.~Berger, B.~Bollob\'as, C.~Borgs, J.T.~Chayes, and O.~Riordan, 
Degree distribution of the FKP network model, 
LNCS {\bf 2719}, 725 (2003). 

\bibitem{bbcsk04}
N.~Berger, C.~Borgs, J.T.~Chayes, R.M.~D'Souza, and R.D.~Kleinberg, 
Competition-induced preferential attachment,  
cond-mat/0402268 (2004). 

\bibitem{yjb01}
S.-H. Yook, H. Jeong, and A.-L. Barab\'asi,  
Modeling the Internet's large-scale topology, 
PNAS {\bf 99}, 13382 (2002). 

\bibitem{a02} 
H.~Agrawal,  
Extreme self-organization in networks constructed from gene expression data, 
Phys. Rev. Lett. {\bf 89}, 268702 (2002). 

\bibitem{rcbh02}
A.F.~Rozenfeld, R.~Cohen, D.~ben-Avraham, and S.~Havlin,   
Scale-free networks on lattices, 
Phys. Rev. Lett. {\bf 89}, 218701 (2002). 

\bibitem{xs02}
R.~Xulvi-Brunet and I.M.~Sokolov, 
Evolving networks with disadvantaged long-range connections, 
Phys. Rev. E {\bf 66}, 026118 (2002). 

\bibitem{cbmr04}
V.~Colizza, J.R.~Banavar, A.~Maritan, and A.~Rinaldo, 
Network structures from selection principles, 
Phys. Rev. Lett. 92, 198701 (2004). 

\bibitem{krl00} 
P.L. Krapivsky, S. Redner, and F. Leyvraz,  
Connectivity of growing random network,  
Phys. Rev. Lett. {\bf 85}, 4629 (2000); 
P.L. Krapivsky and S. Redner (2001),  
Organization of growing random networks,   
Phys. Rev. E {\bf 63}, 066123 (2001). 

\bibitem{dms00}
S.N. Dorogovtsev, J.F.F. Mendes, and A.N. Samukhin,  
Structure of growing networks with preferential linking,  
Phys. Rev. Lett. {\bf 85}, 4633 (2000); 
S.N. Dorogovtsev and J.F.F. Mendes,   
Scaling properties of scale-free evolving networks: Continuum approach, 
Phys. Rev. E {\bf 63}, 056125 (2001). 





\end{thebibliography}
\end{document}